\documentclass[letterpaper, 10 pt, conference]{ieeeconf}  

\IEEEoverridecommandlockouts                              
                                                          
\title{\LARGE \bf
Compositional Construction of Finite State Abstractions\\ for Stochastic Control Systems}

\author{Kaushik Mallik, Sadegh Esmaeil Zadeh Soudjani, Anne-Kathrin Schmuck and Rupak Majumdar
\thanks{All authors are with MPI-SWS, Kaiserslautern, Germany
        {\tt\small \{kmallik, sadegh, akschmuck, rupak\}@mpi-sws.org}}}%

\usepackage{macros}
\usepackage{amssymb,latexsym,amsfonts,amsmath}

\usepackage{eso-pic}
\usepackage{epsfig}
\usepackage{graphicx}
\usepackage{color}

\usepackage{etoolbox}
\usepackage{paralist}
\usepackage{stmaryrd}
\usepackage{xspace}
\usepackage{enumerate}
\usepackage{keyval,trig}
\usepackage{amsmath,amssymb,amsfonts,dsfont}
\usepackage{mathrsfs}

\newtheorem{theorem}{Theorem}[section]
\newtheorem{lemma}[theorem]{Lemma}

\newtheorem{proposition}[theorem]{Proposition}
\newtheorem{corollary}[theorem]{Corollary}

\newtheorem{definition}[theorem]{Definition}

\newtheorem{remark}[theorem]{Remark}
\numberwithin{equation}{section}
\newtheorem{assumption}{Assumption}

\newcommand{\R}{{\mathbb{R}}}
\newcommand{\N}{{\mathbb{N}}}

\DeclareMathOperator{\diff}{d}
\newcommand{\ra}{\rightarrow}
\newcommand{\sigalg}{\mathcal{F}}
\newcommand{\filtration}{\mathbb{F}}
\newcommand{\ul}{\underline}
\newcommand{\ol}{\overline}
\newcommand{\Let}{:=}
\newcommand{\EE}{\mathbb{E}}
\newcommand{\PP}{\mathbb{P}}
\newcommand{\traj}[3]{#1_{#2#3}}
\newcommand{\suprem}{\mathrm{sup}}
\newcommand{\set}[1]{{\{ #1 \}}}
\newcommand{\Sys}[1]{\ifstrempty{#1}{S}{S_{#1}}}
\newcommand{\real}[1]{\ifstrempty{#1}{\mathbb{R}}{\mathbb{R}^{#1}}}
\newcommand{\posreal}[1]{\real{#1}_{>0}}
\newcommand{\poszreal}{{\real{}_{\geq 0}}}
\newcommand{\delimit}{\hspace{1.5mm}}
\newcommand{\norm}[1]{\|\! ~#1~\!\|}
\newcommand{\inorm}[1]{{\|\! ~#1~\!\|}}
\newcommand{\dist}[2]{d(#1,#2)}
\newcommand{\DiscSpace}[2]{[#1]_{#2}}
\newcommand{\KInf}{\mathcal{K}_\infty}
\newcommand{\st}{s.t.\xspace}
\newcommand{\transpose}[1]{[#1]^T}
\newcommand{\defeq}{:=}

\newcommand{\DQnt}[1]{\ifstrempty{#1}{\tilde{\varepsilon}}{\tilde{\varepsilon}_{#1}}}
\newcommand{\vmetric}{\mathbf{e}}

\newcommand{\Def}{Def.\xspace}
\newcommand{\Thm}{Thm.\xspace}
\newcommand{\Lem}{Lemma\xspace}
\newcommand{\Sec}{Sec.\xspace}

\newcommand{\Eqn}{Eqn.\xspace}
\newcommand{\Ass}{Assump.\xspace}

\newcommand{\Prop}{Prop.\xspace}
\newcommand{\Cor}{Cor.\xspace}
\newcommand{\Rem}{Rem.\xspace}

\newcommand{\Lypv}[1]{\ifstrempty{#1}{V}{V_{#1}}}
\newcommand{\gammafunc}[1]{\widehat{\gamma}_{#1}}
\newcommand{\alphalow}[1]{\ifstrempty{#1}{\underline{\alpha}}{\underline{\alpha}_{#1}}}
\newcommand{\alphahigh}[1]{\ifstrempty{#1}{\overline{\alpha}}{\overline{\alpha}_{#1}}}
\newcommand{\sigmad}[1]{\ifstrempty{#1}{\sigma_{d}}{\sigma_{d,#1}}}
\newcommand{\sigmau}[1]{\ifstrempty{#1}{\sigma_{u}}{\sigma_{u,#1}}}
\newcommand{\eqclass}[1]{\mathit{R}_{\varepsilon_{#1}\tilde{\varepsilon}_{#1}}}

\newcommand{\IPECurve}[1]{\ifstrempty{#1}{\mathcal{U}}{\mathcal{U}_{#1}}}
\newcommand{\IPICurve}[1]{\ifstrempty{#1}{\mathcal{W}}{\mathcal{W}_{#1}}}

\newcommand{\ContSysSymb}[1]{\ifstrempty{#1}{\Sigma}{\Sigma_{#1}}}

\newcommand{\Wt}{\widetilde{W}}
\newcommand{\wt}{\tilde{w}}

\newcommand{\IntCon}{\mathcal{I}}
\newcommand{\nbr}{\mathcal{N}}

\newcommand{\State}[1]{\ifstrempty{#1}{X}{X_{#1}}}

\newcommand{\TimeQnt}[1]{\ifstrempty{#1}{\tau}{\tau_{#1}}}
\newcommand{\ContSysDT}[1]{\mathcal{P}_{\TimeQnt{}}(#1)}

\newcommand{\StateDT}[1]{\ifstrempty{#1}{X_{\TimeQnt{}}}{X_{#1,\TimeQnt{}}}}

\newcommand{\IPECurveDT}[1]{\ifstrempty{#1}{\mathcal{U}_\TimeQnt{}}{\mathcal{U}_{#1,\TimeQnt{}}}}
\newcommand{\IPICurveDT}[1]{\ifstrempty{#1}{\mathcal{W}_\TimeQnt{}}{\mathcal{W}_{#1,\TimeQnt{}}}}

\newcommand{\IPETraj}[1]{\ifstrempty{#1}{\mu}{\mu_{#1}}}
\newcommand{\IPITraj}[1]{\ifstrempty{#1}{\nu}{\nu_{#1}}}

\newcommand{\StateQnt}[1]{\ifstrempty{#1}{\eta}{\eta_{#1}}}
\newcommand{\IPQnt}[1]{\ifstrempty{#1}{\omega}{\omega_{#1}}}

\newcommand{\ContSysA}[2]{\ifstrempty{#1}{\mathcal{P}_{\AllQnt{}}(#2)}{\mathcal{P}_{\AllQnt{#1}}(#2)}}
\newcommand{\ContSysAi}[1]{\mathcal{P}_{\AllQnt{i}}(#1)}
\newcommand{\AllQnt}[1]{\ifstrempty{#1}{\TimeQnt{}\StateQnt{}\IPQnt{}}{\TimeQnt{}\StateQnt{#1}\IPQnt{#1}}}

\newcommand{\StateA}[1]{\ifstrempty{#1}{\State{\AllQnt{}}}{\State{#1,\AllQnt{#1}}}}

\newcommand{\IPECurveA}[1]{\ifstrempty{#1}{\IPECurve{\AllQnt{}}}{\IPECurve{#1,\AllQnt{#1}}}}

\newcommand{\IPICurveA}[1]{\ifstrempty{#1}{\IPICurve{\AllQnt{}}}{\IPICurve{#1,\AllQnt{#1}}}}

\newcommand{\deltaA}[1]{\ifstrempty{#1}{\delta_{\AllQnt{}}}{\delta_{#1,\AllQnt{#1}}}}

\newcommand{\AllQntb}[1]{\ifstrempty{#1}{\TimeQnt{}\StateQnt{}\IPQnt{}}{\TimeQnt{}\StateQnt{#1}\IPQnt{#1}}}
\newcommand{\StateAb}[1]{\ifstrempty{#1}{\State{\AllQntb{}}}{\State{#1,\AllQntb{#1}}}}

\begin{document}
\maketitle
\thispagestyle{empty}
\pagestyle{empty}

\begin{abstract}
Controller synthesis techniques for continuous systems with respect to temporal logic specifications typically use a finite-state symbolic abstraction of the system. 
Constructing this abstraction for the entire system is computationally expensive, and does not
exploit natural decompositions of many systems into interacting components. 
We have recently introduced a new relation, called (approximate)
\emph{disturbance bisimulation} for compositional symbolic abstraction 
to help scale controller synthesis for temporal logic to larger systems.

In this paper, we extend the results to stochastic control systems modeled by
stochastic differential equations. 
Given any stochastic control system satisfying a stochastic version of the incremental input-to-state stability property and a positive error bound, we show how to construct a finite-state transition system (if there exists one) which is disturbance bisimilar to the given stochastic control system.
Given a network of stochastic control systems, we give conditions on the simultaneous existence of disturbance bisimilar abstractions to every component allowing for compositional abstraction of the network system.
\end{abstract}

\section{Introduction}
\label{sec:intro}

In abstraction-based controller synthesis, a finite-state symbolic model of a continuous system 
is used to synthesize a symbolic controller for a logical specification, and
the controller is then refined to a controller for the original system. 
This technique has recently gained a lot of attention due to 
two main advantages. 
First, it allows for fully automated controller synthesis for systems with continuous dynamics 
while handling complex specifications (given e.g.\ as $\omega$-regular languages) in 
addition to stability. 
Second, it naturally accounts for the complex interplay between discrete and continuous components within a control loop.
The soundness of the abstraction-based synthesis technique relies on notions of behavioral closeness of 
the original system and its abstraction, which is formalized using system equivalence relations 
(see e.g.\ \cite{TabuadaBook,ReissigWeberRungger_2016} and the references therein). 
Recently, abstraction-based controller synthesis has been extended to stochastic control systems \cite{zamani2014symbolic,jagtap2017automated}.
In the stochastic setting, behavioral closeness of the original system and its abstraction is formalized using the 
$n$-th moment of the trajectories.

Despite its nice theoretical properties and applicability to many different system classes, 
abstraction based controller synthesis does not scale very well because 
both the abstraction step and the controller synthesis step are 
exponential in the dimension of the continuous state space. 
This issue motivated us to propose \emph{disturbance bisimulation} \cite{majumdar2016compositional}, an equivalence relation that exploits
the intrinsic compositionality of systems. 
Given a network of metric systems and their equivalently interconnected disturbance bisimilar abstractions, the main result of 
\cite{majumdar2016compositional} shows that the overall network system is also disturbance bisimilar to the network of the abstractions.  
This result has an interesting consequence: given a construction of a disturbance bisimilar abstraction for 
a given system class, we can compositionally abstract a network of systems from this class 
whenever all local abstractions are guaranteed to exist simultaneously.
In \cite{majumdar2016compositional} we exploited this fact for large networks of incremental input-to-state 
stable \emph{deterministic} control systems and demonstrated the effectiveness of our approach in a case study.

In this paper, we extend these results to the class of 
\emph{stochastic control systems} which satisfy a stochastic version of the incremental input-to-state stability condition.
Our main contribution in this paper is to show how a \emph{stochastic control system}, which 
may be connected to other components within a network, 
allows for the algorithmic computation of 
(1)~a metric system capturing its time-sampled dynamics and 
(2)~a metric system capturing its abstract symbolic dynamics, such that the two constructed systems are disturbance bisimilar. 
This construction allows us to use the results from \cite{majumdar2016compositional} 
to provide a compositional abstraction-based controller synthesis technique for a given network of continuous and stochastic dynamical systems.

Our results relate to recent results in \cite{zamani2015construction} on compositional abstraction for stochastic systems. 
The main difference between our work and \cite{zamani2015construction} is the type of abstraction: 
while we work with finite state symbolic abstractions, their abstractions are infinite state.
There have been some efforts for improving scalability of abstraction techniques for stochastic systems in a different setting, where
abstract models are Markov chains and the goal is to match distributions on states up to a fixed horizon. 
In \cite{DBLP:journals/siamads/SoudjaniA13}, the state space discretization is done adaptively and,
in \cite{DBLP:conf/concur/SoudjaniAM15}, the abstract state space of a monolithic system is represented compositionally.

The paper is organized as follows. 
After introducing preliminaries on stochastic control systems in \Sec~\ref{sec:prelim_stochastic}, 
we define metric systems in \Sec~\ref{existence}, and present how the two particular metric systems discussed above 
can be obtained from a stochastic control system. 
Given these two metric systems, we give sufficient conditions for them to be disturbance bisimilar 
in \Sec~\ref{sec:Dist_bisimulation}, after recalling the notion of disturbance bisimulation from \cite{majumdar2016compositional}. 
In \Sec~\ref{sec:comp_abstraction_stochastic}, we invoke results from \cite{majumdar2016compositional} to 
extend our result from a single stochastic control system to a network of such systems.  All proofs can be found in the appendix.

\section{Stochastsic Control Systems}\label{sec:prelim_stochastic}

Most part of this section is adapted from \cite{zamani2014symbolic} to systems with stochastic disturbance inputs.

\subsection{Notation}

We use the symbols $\N$, $\R$, $\R_{>0}$, $\R_{\geq 0}$ and $\mathbb{Z}$ to denote the set of natural, 
real, positive real, non-negative real numbers, and integers, respectively. 
The symbols $I_n$, $0_n$, and $0_{n\times{m}}$ denote the identity matrix, the zero vector, and the zero matrix 
in $\R^{n\times{n}}$, $\R^n$, and $\R^{n\times{m}}$, respectively. 
Given a vector \mbox{$x\in\mathbb{R}^{n}$}, we denote by $x_{i}$ the $i$-th element of $x$ and
by $\norm{x}$ the infinity norm of $x$. 

A continuous function \mbox{$\gamma:\mathbb{R}_{\geq 0}\rightarrow\mathbb{R}_{\geq 0}$} 
is said to belong to class $\mathcal{K}$ if it is strictly increasing and \mbox{$\gamma(0)=0$}; 
$\gamma$ is said to belong to class $\mathcal{K}_\infty$ 
if \mbox{$\gamma\in\mathcal{K}$} and \mbox{$\lim_{r\rightarrow\infty}\gamma(r)=\infty$}. 
A continuous function \mbox{$\beta:\mathbb{R}_{\geq 0}\times\mathbb{R}_{\geq 0}\rightarrow\mathbb{R}_{\geq 0}$} 
is said to belong to class $\mathcal{KL}$ if, for each fixed $s$, the function \mbox{$\beta(\cdot,s):\mathbb{R}_{\geq 0}\rightarrow\mathbb{R}_{\geq 0}$} 
belongs to class $\mathcal{K}_\infty$ and, for each fixed $r$, the map 
$\beta(r,\cdot):\mathbb{R}_{\geq 0}\rightarrow\mathbb{R}_{\geq 0}$ is decreasing and \mbox{$\lim_{s\rightarrow\infty}\beta(r,s) = 0$}.
Let $f:\mathbb{R}_{\geq 0}\rightarrow\mathbb{R}^k$ be a measurable function.
We define the (essential) supremum $\inorm{f}$ of $f$ as 
$\inorm{f} := \mathrm{ess(sup)}\set{ \norm{f(t)} \mid t\geq 0}$. 
A function $f$ is bounded if $\inorm{f} < \infty$. 
Given a square matrix $M$, we denote by $Tr(M)$ the trace of $M$, and by $\lambda_{min}(M)$ and $\lambda_{max}(M)$ the minimum and maximum eigenvalue of $M$ respectively. Given a matrix \mbox{$M=\{m_{ij}\}\in\R^{n\times{m}}$}, we denote by $\norm{M}:=\max_{1\leq{i}\leq{n}}\sum_{j=1}^m\vert{m_{ij}\vert}$ the infinity 
norm of $M$, 
and by $\|M\|_F \Let \sqrt{\text{Tr}\left(MM^T\right)}$ the Frobenius norm of $M$.
We denote by
$Diag(a_1,\ldots,a_n)$ the diagonal matrix with diagonal entries $a_1,\ldots,a_n\in \R$.
If $a_1,\ldots,a_n$ are matrices, then $Diag(a_1,\ldots,a_n)$ is a block diagonal matrix of appropriate dimension.

\subsection{Stochastic Control System}
We fix the probability space for the whole paper as $(\Omega, \sigalg, \PP)$, where $\Omega$ is a sample space, $\sigalg$ is a sigma algebra over $\Omega$ representing the set of events, and $\PP$ is a probability measure. Let $(\Omega, \sigalg, \PP)$ admits a filtration $\filtration = (\sigalg_s)_{s\geq 0}$ which is complete and right continuous \cite[p.\ 89]{ref:KarShr-91}. Let $(B_s)_{s \ge 0}$ be a $r$-dimensional $\filtration$-Brownian motion.

\begin{definition} \label{Def_control_sys}
A stochastic control system is a tuple $\Sigma=(X,U,\mathcal{U},W,\mathcal{W},f,\sigma)$, where
$X = \mathbb{R}^{n}$ is the state space, 
$U\subseteq\R^m$ is the input set that is assumed to be compact, 
$\mathcal{U}$ is a subset of set of all measurable, locally essentially bounded functions of time from $\R_{\geq 0}$ to $U$, 
$W\subseteq\R^p$ is the disturbance input space that is assumed to be compact, 
$\mathcal{W}$ is 
a set of stochastic processes with elements \mbox{$\nu:\Omega\times\R_{\geq 0}\rightarrow W$}, 
$f:X\times U \times W \rightarrow X$ is a continuous function of its arguments representing the \emph{drift} of $\Sigma$, 
 $\sigma:X\rightarrow\R^{n\times{r}}$ is a function representing the \emph{diffusion} of $\Sigma$.
\end{definition}

A stochastic process \mbox{$\xi:\Omega \times \R_{\geq 0} \rightarrow X$} is called a \textit{solution process} of $\Sigma$ if there exists $\mu\in\mathcal{U}$ and $\nu\in \mathcal{W}$ satisfying the following stochastic differential equation:
\begin{equation}
\label{eq0}
	\diff \xi(t)= f(\xi(t),\mu(t),\nu(t))\diff t+\sigma(\xi(t))\diff B_t,
\end{equation}
 $\PP$-almost surely ($\PP$-a.s.).  
 For succinctness of representation, we use the notation $\xi_{a \mu\nu}$ to denote a stochastic solution process of $\Sigma$ from the initial condition $\xi_{a \mu\nu}(0) = a$ $\PP$-a.s., and under effect of input signal $\mu\in \mathcal{U}$ and disturbance signal $\nu\in\mathcal{W}$. Note that given any time instant $t$,  $\xi_{a \mu\nu}(t)$ represents a random variable from $\Omega$ to $X$ measurable in $\sigalg_t$.

We make the following two assumptions on stochastic control systems to ensure a unique global continuous solution.
\begin{assumption}[Lipschitz condition]
	\label{ass:Lip}
	There exist constants $L_f,L_\sigma\in\R_{\ge 0}$ such that the following inequalities hold
	$\Vert f(x,u,w)-f(x',u',w')\Vert \leq L_f(\Vert x-x'\Vert  +  \Vert u-u'\Vert  + \Vert w-w'\Vert) $,
	and
	$\Vert\sigma(x)-\sigma(x')\Vert \leq L_\sigma\Vert{x}-x'\Vert $,
	for all $x,x'\in X$, $u,u'\in U $ and $w,w'\in W$.
\end{assumption}

\begin{assumption}[Linear growth]
		\label{ass:linear_growth}
	There exists a positive constant $K$ such that for all $x\in X,\,u,\in U $ and $w\in W$,
	\begin{equation}
	\label{eq:lin_growth}
		\max(\|f(x,u,w)\|^2,\|\sigma(x)\|^2)\le K\left(1+\|x\|^2\right).
	\end{equation}
\end{assumption}
\Ass~\ref{ass:Lip} on Lipschitz continuity gives uniqueness and \Ass~\ref{ass:linear_growth} on linear growth gives global existence (\cite[\Thm~5.2.1]{oksendal}).
The latter will also be used in \Sec~\ref{sec:CA:Assumptions} (cf.~\Prop~\ref{prop:bound on moment}) to provide an upper bound on the second moment of the solution process.

In this paper the disturbances in the set $\mathcal W$ are allowed to be stochastic. This is necessary because, as will be described later, in our setting the disturbances play the role of trajectories of other stochastic control systems after interconnection. For the results of this paper to hold, we require the process $(\xi,\nu):\Omega\times \R_{\ge 0}\rightarrow X\times W$ to be an It\^{o} process, i.e., $(\xi,\nu)$ has to be the solution of a possibly time-inhomogeneous It\^{o} diffusion.

\subsection{$\delta$-ISS-$M_q$}
\label{sec3}

We now generalize the notion of incremental input-to-state stability in the $q$-the moment
($\delta$-ISS-$M_q$) for stochastic control
systems from \cite{zamani2014symbolic} by considering disturbances.
In the absence of noise, these notions correspond to $\delta$-ISS for deterministic systems \cite{angeli}. 

\begin{definition} 
\label{dISS}
A stochastic control system $\Sigma=(X, U ,\mathcal{U},W,\mathcal{W},f,\sigma)$ is stochastically incrementally input-to-state stable in the $q$-th moment ($\delta$-ISS-$M_q$), if there exists a $\mathcal{KL}$ function $\beta$ and $\KInf $ functions $\rho_u$ and $\rho_d$ such that for any $t\in{\mathbb{R}_{\geq 0}}$, any $\mu$, ${\mu}'\in\mathcal{U}$, any $\nu$, $\nu'\in\mathcal W$, and any \mbox{$\R^n$-valued} random variables $a$ and $a'$ that are measurable in $\sigalg_0$, the following condition is satisfied:
\begin{multline}
\EE \left[\left\Vert  \xi_{a\mu\nu}(t)-\xi_{a'{\mu}'\nu'}(t)\right\Vert ^q\right] \leq\beta\left( \EE\left[ \left\Vert a-a' \right\Vert ^q\right], t \right)\\
 + \rho_u \left(\inorm{{\mu} - {\mu}'} \right) + \rho_d \left(\EE\left[\inorm{{\nu} - {\nu}'}^q\right] \right). \label{delta_PISS}
\end{multline}
\end{definition}

The $\delta$-ISS-$M_q$ property can be characterized in terms of the existence of {\em stochastic incremental Lyapnuov functions}.

\begin{definition} \cite[\Def~3.2]{zamani2014symbolic}
\label{delta_PISS_Lya}
Define the \emph{diagonal set} $\Delta$ as 
\mbox{$\Delta=\left\{(x,x) \mid x\in \R^n\right\}$}.
Consider a stochastic control system $\Sigma=(X, U ,\mathcal{U},W,\mathcal{W},f,\sigma)$ 
and a continuous function $V:X\times X\rightarrow\mathbb{R}_{\geq 0}$ which is 
smooth on
$\{\R^n\times\R^n\}\backslash\Delta$. 
Function $V$ is called a 
$\delta$-ISS-$M_q$ \emph{Lyapunov function} for $\Sigma$ 
if there exist $\KInf $ functions 
$\underline{\alpha}$, $\overline{\alpha}$, $\sigma_u$, $\sigma_d$, and a constant $\kappa\in\mathbb{R}_{> 0}$ such that
\begin{itemize}
\item[(i)] $\ul{\alpha}$ is a convex function, and $\ol \alpha$ and $\sigma_d$ are concave functions;
\item[(ii)] for any $x,x'\in X$,\\ 
$\underline{\alpha}(\Vert x-x'\Vert ^q)\leq{V}(x,x')\leq\overline{\alpha}(\Vert x-x'\Vert ^q)$;
\item[(iii)] for any $x,x'\in X$, $x\neq x'$, any $u,u'\in U $, and any $w$,$w'\in W$,
\begin{align*}
	&\mathcal{L}^{u,u',w,w'} V(x, x') 
	 \Let \left[\partial_xV~~\partial_{x'}V\right] \begin{bmatrix} f(x,u,w)\\f(x',u',w')\end{bmatrix}\\\notag
	 &+ \frac{1}{2} \text{Tr} \left(\begin{bmatrix} \sigma(x) \\ \sigma(x') \end{bmatrix}\left[\sigma^T(x)~~\sigma^T(x')\right] \begin{bmatrix}
\partial_{x,x} V & \partial_{x,x'} V \\ \partial_{x',x} V & \partial_{x',x'} V
\end{bmatrix}	\right) \\
	& \leq-\kappa V(x,x')+\sigma_u(\norm{u-u'})+\sigma_d(\norm{w-w'}^q),
\end{align*} 
\end{itemize}
where $\mathcal{L}^{u,u',w,w'}$ is the infinitesimal generator (\cite[Section 7.3]{oksendal}) associated to the stochastic control system \eqref{eq0}, 
which depends on two separate controls $u, u'\in U$ and two separate disturbances $w,w'\in W$.  
In this case we say that the stochastic control system $\Sigma$ \emph{admits} a $\delta$-ISS-$M_q$ Lyapunov function, 
\emph{witnessed} by $\underline{\alpha}$, $\overline{\alpha}$, $\sigma_u$, $\sigma_d$, and $\kappa\in\mathbb{R}_{> 0}$.
\end{definition}

Note that condition $(i)$ is not required in the context of deterministic control systems. 
Condition $(ii)$ implies that the growth rates of the functions $\ol{\alpha}$ and $\ul{\alpha}$ are linear, as a concave function is supposed to dominate a convex 
one. 
These conditions are not restrictive provided we are interested in the dynamics of $\Sigma$ on a compact subset $D\subset\R^n$, which is often the case in practice. 
It can be readily verified that the $\delta$-ISS-$M_q$ Lyapunov function in Definition \ref{delta_PISS_Lya} 
is a stochastic bisimulation function between $\Sigma$ and itself, as defined in \cite{julius1}, Def.~5.

The following theorem 
describes $\delta$-ISS-$M_q$ in terms of the existence of $\delta$-ISS-$M_q$ Lyapunov functions.
It generalizes the corresponding theorem \cite[\Thm~3.3]{zamani2014symbolic} in the presence of disturbances.

\begin{theorem}
\label{the_Lya}
A stochastic control system $\Sigma$ is $\delta$-ISS-$M_q$ if it admits a $\delta$-ISS-$M_q$ Lyapunov function. 
\end{theorem}

In order to keep the notation simple, we present the results only for \emph{second moment} 
in the rest of paper with the understanding that they can be generalized for other moments.

The following lemma (compare \cite[\Lem~3.4]{zamani2014symbolic}) provides a sufficient condition on a particular 
function $V$ to be a $\delta$-ISS-$M_q$ Lyapunov function.

\begin{lemma}\label{lem:lyapunov}
	Let $\Sigma=(X, U ,\mathcal{U},W,\mathcal{W},f,\sigma)$ be a stochastic control system. 
	Let $P\in\R^{n\times{n}}$ be a symmetric positive definite matrix.
        Consider the function $V: \R^n \times \R^n \ra \R_{\geq 0}$ defined as: 
	\begin{align}
	\label{V}
			V(x,x')\Let\frac{1}{2}(x-x')^TP(x-x') 
	\end{align}
        and satisfying
	\begin{align}
	\label{nonlinear ineq cond}
	(x-x')^T P(f(x,u,w)-f(x',u,w))&+\\
	+\frac{1}{2} \left \| \sqrt{P} \left( \sigma(x) - \sigma(x')\right) \right \|_F^2&\le-2\widetilde{\kappa}V(x,x'),\nonumber
	\end{align}
	or, if $f$ is differentiable, satisfying
	\begin{align}
	\label{nonlinear ineq cond1}
		(x-x')^T P \partial_x f(z,u,w)(x-x') +\quad& \\\nonumber
		+\frac{1}{2} \left \| \sqrt{P} \left( \sigma(x) - \sigma(x')\right) \right \|_F^2&\le-2\widetilde{\kappa} V(x,x'),	
	\end{align}
for all $x,x',z\in X$, for all $u \in  U $, for all $w\in W$, and for some constant $\widetilde{\kappa}\in\R_{> 0}$. 
Then $V$ is a $\delta$-ISS-$M_2$ Lyapunov function for $\Sigma$.
\end{lemma}

\subsection{Noisy and Noise Free Trajectories}

In this section we provide an upper bound on the distance between a stochastic state trajectory and its associated noise-free 
trajectory at any instant of time. 
This bound is a generalization of the bound in \cite[\Lem~3.10]{zamani2014symbolic} to the case when there is disturbance in the system. 
The bound will be instrumental in proving closeness between the trajectories of a stochastic control system and its abstraction in \Sec~\ref{existence}.

\begin{lemma}
\label{lem:moment est}
	Consider a stochastic control system $\Sigma=(X, U ,\mathcal{U},W,\mathcal{W},f,\sigma)$. Suppose there exists a $\delta$-ISS-$M_2$ Lyapunov function $V$ of $\Sigma$ s.t. its Hessian matrix in $\R^{2n\times 2n}$ satisfies $0\leq \partial_{x,x}V(x,x')\leq P$, for some positive semi-definite matrix $P\in \R^{2n\times 2n}$ and for any $x,x'\in\R^{n}$. Define $\ol \xi_{x\mu\nu}$ as the solution of the ordinary differential equation (ODE)
	\begin{equation}
	\label{eq:ODE}
		\traj{\dot{\ol \xi}}{x}{\mu\nu}(t)=f\left(\traj{\ol \xi}{x}{\mu\nu}(t),\mu(t),\nu(t)\right)
	\end{equation}
	starting from the initial condition $x$.
	Then for any $x$ in a compact set $D\subset\R^n$, any $\mu\in\mathcal{U}$ and any $\nu\in\mathcal{W}$, we have 
	\begin{equation*}
	 	\EE \left[\left\Vert\traj{\xi}{x}{\mu\nu}(t)-\traj{\ol \xi}{x}{\mu\nu}(t)\right\Vert^2\right] \le h(\sigma,t),	 
	\end{equation*}
	where
	\begin{align*}
		&h(\sigma,t) := \alphalow{}^{-1}\Big( \frac{1}{2}\norm{\sqrt{P}}^2\cdot n\cdot\mathrm{min}\set{n,r}\cdot\mathrm{e}^{-\kappa t}\cdot L_\sigma^2\cdot\\
		&\qquad\cdot\int_0^t \left[ \beta\left(\suprem_{x\in D}\norm{x}^2,s\right) +\rho_u\left( \suprem_{u\in U}\norm{u}\right)\right.\\
		&\qquad\qquad\left. + \rho_d\left( \suprem_{w\in W}\norm{w}^2 \right) \right]ds\Big). \numberthis \label{eqn h}
	\end{align*}
	 The non-negative valued function $h$ tends to zero as $t \ra 0$, $t\ra \infty$, or as $L_\sigma\ra 0$, where $L_\sigma$ is the Lipschitz constant introduced in
	 \Ass~\ref{ass:Lip}.	
\end{lemma}

\begin{remark}
	\Eqn~\eqref{eqn h} gives a representation of the function $h$ in terms of $\rho_u$ and $\rho_d$. This representation of $h$ can be translated into a form using $\sigma_u$ and $\sigma_d$ instead, as  shown in \cite[\Lem ~3.10, \Cor~3.11]{zamani2014symbolic}. 
\end{remark}

\section{From Stochastic Control Systems to Metric Systems}
\label{existence}

We now introduce (deterministic) metric systems and interpret 
stochastic control systems and their abstractions as metric systems. 
As in \cite{zamani2014symbolic,majumdar2016compositional}, 
we consider metric systems that are time sampled w.r.t.\ 
a globally fixed time sampling parameter $\tau \in \posreal{}$. 

\begin{definition}\label{def:metric system}
Given the probability space $\Omega$ and a time sampling parameter $\tau\in \posreal{}$,
 a \emph{stochastic metric system}\footnote{
Often, metric systems are defined with an additional output space and an output map from states to the output space.
We omit the output space for notational simplicity; for us, the state
and the output space coincide, and the output map is the identity function.}
$S=(X,U,\mathcal{U}_\tau,W,\mathcal{W}_\tau,\delta_\tau)$
consists of
a (possibly infinite) set of states 
$X$, given by a set of random variables, and equipped with a metric $d:X\times X \rightarrow \poszreal$,
a set of piece-wise constant inputs $\mathcal{U}_\tau$ of duration $\tau$ taking values in $U\subseteq \real{m}$, i.e.,
\begin{equation}
\label{equ:def:mathcalU}
	\mathcal{U}_\tau = \set{
		 \mu:[0,\TimeQnt{}]\rightarrow U \mid
		 \forall t_1,t_2\in[0,\TimeQnt{}]~.~\mu(t_1) = \mu(t_2)},
\end{equation}
a set of disturbances $\mathcal{W}_\tau$ taking values in $W\subseteq \real{p}$, i.e.,
\begin{equation}
\label{equ:def:mathcalW}
	\mathcal{W}_\tau \subseteq \set{
		\nu:\Omega \times [0,\TimeQnt{}]\rightarrow W},
\end{equation}
and a transition function $\delta_\tau: X\times \mathcal{U}_\tau \times \mathcal{W}_\tau \rightarrow 2^{X}$. We write $x\xrightarrow[\tau]{\mu, \nu} x'$ if $x'\in\delta_\tau(x,\mu, \nu)$, and we denote the unique value of $\mu\in\mathcal{U}$ over $[0,\tau]$ by  $u_\mu\in U$.
\end{definition}

A \emph{deterministic metric system} is a special type of a stochastic metric system where the states are deterministic points (i.e. random variables with Dirac delta distributions), and disturbances are deterministic signals of the form $[0,\tau]\rightarrow W$.

If the metric system $S$ is undisturbed, we define $W=\set{0}$. In this case we occasionally represent $S$ by the tuple $S = (X, U, \mathcal{U}_\tau, \delta_\tau)$ and use $\delta_\tau:X\times \mathcal{U}_\tau\rightarrow 2^X$ with the understanding that $x'\in\delta_\tau(x,\mu, \nu)$ holds for the zero trajectory $\IPITraj{}:{\poszreal}\rightarrow \set{0}$ whenever $x'\in\delta_\tau(x,\mu)$.
By slightly abusing notation we write $x'=\delta_{\tau}(x,\mu, \nu)$ as a short form when the set $\delta_{\tau}(x,\mu, \nu) = \set{x'}$ is a singleton.
If $X$, $\mathcal{U}_{\tau}$ and $\mathcal{W}_{\tau}$ are finite (resp. countable), $S$ is called \emph{finite} (resp. \emph{countable}). 
We also assign to a transition $x'=\delta_{\tau}(x,\mu, \nu)$  any continuous time evolution $\xi:[0,\tau]\rightarrow X$ \st $\xi(0) = x$ and $\xi(\tau) = x'$.

In the following we introduce two approaches to capture an abstracted version of the dynamics of a stochastic control systems $\Sigma$ by a metric system conforming to \Def~\ref{def:metric system}. The first approach results in a sampled time abstraction which we denote by $\ContSysDT{\Sigma}$.

\begin{definition}
	\label{def:control-sys-disctime}
Given a stochastic control system $\Sigma = (X, U, \mathcal{U}, W, \mathcal{W}, f,\sigma)$,   
a \emph{time-sampling parameter} $\tau \in \posreal{}$, 
and a probability space $(\Omega, \sigalg, \PP )$, 
the
\emph{discrete-time stochastic metric system} induced by $\Sigma$ is defined by
$\ContSysDT{\Sigma} = (X_{\mathsf r}, U, \mathcal{U}_\tau, W, \mathcal{W}_\tau, \delta_\tau)$
s.t. 
$X_{\mathsf r}$ is the set of all $X$-valued random variables,  
$\mathcal{U}_{\tau}$ and $\mathcal{W}_{\tau}$ are defined over $U$ and $W$, respectively, as in \Eqn ~\eqref{equ:def:mathcalU}-\eqref{equ:def:mathcalW} and
$\delta_\tau(x,\mu,\nu) = x'$ if $x$ and $x'$ are measurable in $\mathcal{F}_t$ and $\mathcal{F}_{t+\tau}$, respectively, for some $t\in \R_{\ge 0}$, and there exists a solution process $\xi:\Omega\times\R_{\ge 0}\rightarrow \mathbb{R}^n$ of $\Sigma$ satisfying $\xi(0) = x$ and $\xi_{x\mu\nu}(\tau) = x'$ $\mathbb{P}$-a.s. 
Since we allow any state to be initial, all states in $X$ need to be measurable on $\mathcal{F}_0$. 
We equip $X_{\mathsf r}$ with the metric $\dist{x}{x'} \defeq \left( \mathbb{E}[\norm{x-x'}^2] \right)^{1/2}$.
\end{definition}

\smallskip

\begin{remark}
Recall that the disturbances in the set $\mathcal{W}$ are stochastic. 
Hence the above metric system must be constructed by looking at the sampled version of $(\xi,\nu)$, which is the solution process of the It\^{o} diffusion associated with $(x,w)$.
\end{remark}

The second approach additionally imposes a quantization of the state, input and disturbance spaces and results in a metric system denoted by $\ContSysA{}{\Sigma{}}$. Before defining this system formally we introduce notation for quantization.
For any $A\subseteq \real{n}$ and any vector $\StateQnt{}$ with elements $\StateQnt{i}>0$, we define 
$\DiscSpace{A}{\StateQnt{}}:=
\set{ (a_1,\ldots, a_n) \in A \mid a_i = 2k\StateQnt{i}, k\in\mathbb{Z}, i=1,\ldots, n}.$
For $x\in\real{n}$ and vector $\lambda$ with elements $\lambda_i>0$, let
$\mathbb{B}_\lambda(x) = \set{x'\in\real{n} \mid \norm{x_i-x_i'}\leq \lambda_i}$ denote the closed rectangle centered at $x$.
Note that for any $\lambda \geq \StateQnt{}$ (element-wise), the collection of sets
$\mathbb{B}_\lambda(q)$ with $q\in \DiscSpace{\real{n}}{\StateQnt{}}$ is a \emph{cover} of $\real{n}$,
that is, $\real{n} \subseteq \cup \set{\mathbb{B}_\lambda(q)\mid q\in \DiscSpace{\real{n}}{\StateQnt{}} }$. We will use this insight to discretize the state and the input space of $\Sigma$ using discretization parameters $\eta$ and $\omega$, respectively.

Also we need to define a vector-valued metric for comparing two disturbance vectors. 
Let $A_1,\ldots, A_k$ be a finite set of metric spaces, where each $A_i$, $i=1,\ldots,k$ has a metric
$d_i:A_i\times A_i \rightarrow \mathbb{R}_{\geq 0}$. 
Let $A = \prod_{i=1}^k A_i$. 
We construct the metric $\vmetric:A\times A \rightarrow \mathbb{R}_{\geq 0}^k$ 
as an extension of  the metrics $d_i$ on $A_i$:
for any $a=(a_1,\ldots, a_k)$ and $a'=(a'_1,\ldots,a'_k)$,
we define
\begin{equation}
\label{equ:defvmetric_composed}
\vmetric(a,a'):= (d_1(a_1,a_1'), \ldots, d_k(a_k, a_k')).
\end{equation}

For the disturbance space $W$ we allow the discretization of $W$ to be predefined. 
We make the following general assumptions on the discretizaion of $W$ which will be useful when we 
deal with networks of stochastic control systems in \Sec~\ref{sec:comp_abstraction_stochastic}.

\begin{assumption}\label{ass:Wt}
Let $\Sigma = (X, U, \mathcal{U}, W, \mathcal{W}, f,\sigma)$ be a stochastic control system.
Then we assume that there exists a countable set $\Wt\subseteq W$, that exists a vector $\tilde{\varepsilon} \in \mathbb{R}_{\geq 0}^{p}$, and
a vector-valued metric $\vmetric:W\times W\rightarrow\mathbb{R}_{\geq 0}^{p}$, 
s.t.\ for all $w\in W$ there exists a $\wt\in\Wt$ for which 
       \begin{equation} \label{equ:Vtildeeps}
        \vmetric(w,\wt)\leq\tilde{\varepsilon} \quad
        \text{and}
        \quad \norm{w-\wt}\leq\norm{\vmetric(w,\wt)}.
       \end{equation}
\end{assumption}
Using this assumption we formally define the abstract metric system $\ContSysA{}{\Sigma}$ induced by $\Sigma$ as follows. 

\begin{definition} \label{def:abstract system}
       Let $\Sigma = (X, U, \mathcal{U}, W, \mathcal{W}, f,\sigma)$ be a stochastic control system for which \Ass~\ref{ass:Wt} holds.  
       Given three constants $\tau \in \posreal{}$, $\StateQnt{} \in \posreal{}$, and $\IPQnt{} \in \posreal{}$, the 
       \emph{discrete-time discrete-space deterministic metric system}
       induced by $\Sigma$ is defined by       	
       	\begin{equation}
       	 \ContSysA{}{\Sigma{}} = (\StateA{}, 
	                                \DiscSpace{U}{\IPQnt{}},\IPECurveA{},
	                                \Wt,\IPICurveA{},
	                                \deltaA{})
       	\end{equation}        	
    \st $\StateA{} = \DiscSpace{X}{\StateQnt{}}$, $\IPECurveA{}$ is defined over $\DiscSpace{U}{\IPQnt{}}$, as in \eqref{equ:def:mathcalU},
       \begin{equation*}\label{equ:piecewise constant W}
       		\IPICurveA{} := \set{\nu:[0,\tau]\rightarrow \Wt \mid \forall t,k\in [0,\tau]~.~\nu(t) = \nu(k)},
       \end{equation*}
       and 
		 \begin{align*}
		  &\deltaA{}(x, \mu, \nu) = \set{x'\in\StateAb{} \mid \norm{\traj{\ol \xi}{x}{\mu\nu}(\tau) - x'} \leq \StateQnt{}},
		 \end{align*}
		 where $\traj{\ol \xi}{x}{\mu\nu}(\cdot)$ are the noise free trajectories of $\Sigma$ defined via \Eqn~\eqref{eq:ODE}.
	We equip $\StateA{}$ with the metric \mbox{$\dist{x}{x'} \defeq \norm{x -x'}$} naturally inherited from $X$.  We denote the unique value of $\nu\in\IPICurveA{}$ over $[0,\tau]$ by  $w_\nu\in \Wt$.
\end{definition}

\begin{remark}
Let us emphasize that even though
$\ContSysDT{\Sigma}$ is a stochastic metric system and $\ContSysA{}{\Sigma}$ is a deterministic metric system, 
since we are interested in studying the closeness of their trajectories in the next section, 
it is important that $X_{\mathsf r}$ and $\StateA{}$ are part of the same state space. We interpret $\StateA{}$ to be contained in $X_{\mathsf r}$, since a set of points can be associated with a set of random variables with Dirac delta distributions.
\end{remark}

\section{Disturbance Bisimulation}
\label{sec:Dist_bisimulation}
This section contains the main contribution of the paper; after recalling the notion of \emph{disturbance bisimulation} from \cite{majumdar2016compositional}
we present sufficient conditions under which the two metric systems $\ContSysA{}{\Sigma}$ and $\ContSysDT{\Sigma}$ associated with a stochastic control system $\Sigma$
are disturbance bisimilar. 
For this analysis, we restrict our attention to $\delta$-ISS-$M_2$ stochastic control systems with $f(0_n,0_m,0_p) =0_n$ and $\sigma(0_n) = 0_{n\times{r}}$, whose evolution is restricted to a compact region $D\subset \mathbb{R}^n$.

\begin{definition} \label{def:DisturbanceBisimulation}
Let $\Sys{i}=(X_i,U_i,\mathcal{U}_{\tau,i},W_i,\mathcal{W}_{\tau,i},\delta_{\tau,i})$, $i=1,2$, be two metric systems, with state-spaces $\State{1},\State{2}\subseteq\State{}$
and disturbance sets
$W_{1},W_{2}\subseteq W\subseteq \mathbb{R}^p$.
Furthermore, let $\State{}$ admit the metric $d:\State{}\times\State{}\rightarrow\mathbb{R}_{\geq 0}$ and $W$ admit the vector-valued metric
$\vmetric:W\times W\rightarrow\mathbb{R}_{\geq 0}^p$.
A binary relation $R \subseteq X_{1}\times X_{2}$ is a \emph{disturbance bisimulation with parameters $(\varepsilon,\tilde{\varepsilon})$} where $\varepsilon\in \poszreal{}$ and $\tilde{\varepsilon}\in\mathbb{R}_{\geq 0}^p$,
iff for each $(x_{1},x_{2})\in R$:
	\begin{enumerate}[(a)]
\item $d(x_{1},x_{2})\leq\varepsilon$; 
\item for every $\IPETraj{1}\in \mathcal{U}_{\tau,1}$ there exists a $\IPETraj{2} \in \mathcal{U}_{\tau,2}$ such that for all $\IPITraj{2}\in \mathcal{W}_{\tau,2}$ and $\IPITraj{1} \in \mathcal{W}_{\tau,1}$ with
$\vmetric(\nu_1(0),\nu_2(0))\leq \tilde{\varepsilon}$, we have that
$(\delta_{\tau,1}(x_{1},\IPETraj{1},\IPITraj{1}), \delta_{\tau,2}(x_{2}, \IPETraj{2}, \IPITraj{2}))\in R$; and
\item for every $\IPETraj{2} \in \mathcal{U}_{\tau,2}$ there exists a $\IPETraj{1}\in \mathcal{U}_{\tau,1}$ such that for all $\IPITraj{1}\in \mathcal{W}_{\tau,1}$ and $\IPITraj{2} \in \mathcal{W}_{\tau,2}$ with
$\vmetric(\nu_1(0),\nu_2(0))\leq \tilde{\varepsilon}$, we have that
$(\delta_{\tau,1}(x_{1},\IPETraj{1},\IPITraj{1}), \delta_{\tau,2}(x_{2}, \IPETraj{2}, \IPITraj{2}))\in R$.
	\end{enumerate}
$\Sys{1}$ and $\Sys{2}$ are said to be \emph{disturbance bisimilar} with parameters $(\varepsilon,\tilde{\varepsilon})$ if
there is a disturbance bisimulation relation 
$R$ between $\Sys{1}$ and $\Sys{2}$ with parameters $(\varepsilon,\tilde{\varepsilon})$.
\end{definition}

In order to prove the existence of a disturbance bisimulation between $\ContSysA{}{\Sigma}$ and $\ContSysDT{\Sigma}$ we require two additional assumptions.  
\begin{assumption} \label{assume:V with triangular inq}
Let $\Sigma$ be a stochastic control system admitting a $\delta$-ISS-$M_2$ Lyapunov function $V$.
There exists a $\mathcal{K}_\infty$ and concave function $\widehat\gamma$ s.t. for any $x,x',x''\in X$,
\begin{equation}\label{supplement}
|V(x,x')-V(x,x'')|\leq\widehat\gamma(\norm{x'-x''}).
\end{equation}
\end{assumption}

This assumption is not restrictive as we are interested in the dynamics of $\Sigma$ on a compact subset $D\subset\R^n$.
\begin{assumption}\label{ass:psi}
 Let $\Sigma$ be a stochastic control system with the associated metric systems $\ContSysDT{\Sigma}$ and $\ContSysA{}{\Sigma}$ introduced in \Sec~\ref{existence}. 
	Then there exists a $\KInf$ function $\psi$ \st for all disturbance pairs $\nu\in\mathcal{W}_{\tau}$ and $\hat{\nu}\in\mathcal{W}_{\tau\eta\omega}$ with $d(\nu(0),\hat{\nu}(0)) \leq  \norm{\tilde{\varepsilon}}$, the following holds for all $t\in[0,\tau]$:
	\begin{align}
		d(\nu(t),\hat{\nu}(t)) = \left(\EE[\norm{\hat{\nu}(t)-\nu(t)}^2]\right)^\frac{1}{2} \leq \psi(t) + \norm{\tilde{\varepsilon}}.
	\end{align}
\end{assumption}

Given \Ass~\ref{assume:V with triangular inq} and \Ass~\ref{ass:psi}, we present our first main result in the following theorem.
\begin{theorem} \label{thm:sim-approx-single}
	Let $\Sigma$ be a stochastic control system admitting a $\delta$-ISS-$M_2$ Lyapunov function 
	${\Lypv{}}$ witnessed by $\kappa$, $\alphalow{}$, $\alphahigh{}$, $\sigmau{}$, and $\sigmad{}$, that satisfies \Ass~\ref{assume:V with triangular inq} with $\KInf$ function $\gammafunc{}$.
	Fix $\TimeQnt{}>0$ and $\Wt\subseteq W$ s.t. \eqref{equ:Vtildeeps} holds and let $\ContSysA{}{\Sigma{}}$ be the countable deterministic metric system associated wih $\Sigma$ according to \Def~\ref{def:abstract system} such that \Ass~\ref{ass:psi} holds. 
If
	\begin{multline}
		0\leq \StateQnt{} \leq \textrm{min}\Big\lbrace(\alphahigh{})^{-1}\circ\alphalow{}(\varepsilon^2),\gammafunc{}^{-1}\big[ (1-\mathsf{e}^{-\kappa\tau})\alphalow{}(\varepsilon^2)\\
		- \frac{1}{\mathsf{e}\kappa}\sigma_u(\omega) - \frac{1}{\mathsf{e}\kappa}\sigma_d(\psi(\tau) + \norm{\tilde{\varepsilon}}) \big] - (h(\sigma,\tau))^{\frac{1}{2}}\Big\rbrace, \label{eq:condition on StateQnt}
	\end{multline}
	where $h(\sigma,\tau)$ is as in \eqref{eqn h},
	then the relation
	\begin{align}
			\eqclass{} = &\left\lbrace (\hat{x},x)\in\StateA{}\times X_{\mathsf r} \ |\ \mathbb{E}[\Lypv{}(\hat{x},x)]\leq\alphalow{}(\varepsilon_{}^2) \right\rbrace \label{eqn:local simulation relation}
	\end{align}
	is a disturbance bisimulation relation (in the second moment) with parameters $(\varepsilon,\tilde{\varepsilon})$ between $\ContSysA{}{\ContSysSymb{}}$ and $\ContSysDT{\ContSysSymb{}}$.
\end{theorem}

\begin{remark}
	Given any fixed $\tau$ and $\tilde{\varepsilon}$, one can always find sufficiently small $\eta$ and $\omega$ s.t. \eqref{eq:condition on StateQnt} and \eqref{eqn:local simulation relation} hold, as long as $\varepsilon$ is lower bounded according to
	\begin{equation}
		\varepsilon^2 > \alphalow{}^{-1}\left( \frac{
		\frac{1}{\mathsf{e}\kappa} \sigmad{}(\psi(\tau) + \norm{\tilde{\varepsilon}}) + \widehat{\gamma}\left( h(\sigma,\tau)^{\frac{1}{2}} \right)}{
		\left( 1- \mathsf{e}^{-\kappa\tau} \right)} \right).
\label{eqn: bound on eps}
	\end{equation}
	The lower bound on $\varepsilon$ can be minimized by choosing an optimal Lyapunov function $V$ for a given system $\Sigma$ (see e.g. \cite[\Rem~3.6]{zamani2014symbolic}). Note that, when the system does not experience any disturbance,  \eqref{eqn: bound on eps} reduces to \cite[V.5]{zamani2014symbolic}.
\end{remark}

\section{Compositional Abstraction}\label{sec:comp_abstraction_stochastic}
Let us first summarize what we have presented so far. In \Sec~\ref{existence} we have introduced two different metric systems $\ContSysDT{\Sigma}$ and $\ContSysA{}{\Sigma{}}$ associated with a given stochastic control system $\Sigma$. Recall that $\ContSysDT{\Sigma}$ is an infinite state system, whereas $\ContSysA{}{\Sigma{}}$ is a finite state system under the assumption that the state space of $\ContSysA{}{\Sigma{}}$ is restricted to a compact subset of $\mathbb{R}^n$. Then we gave sufficient conditions for these two abstractions to be disturbance bisimilar in \Sec~\ref{sec:Dist_bisimulation}.

In this section, we consider a \emph{network} of stochastic control systems $\set{\Sigma_i}_{i\in I}$, and the respective \emph{local} abstractions $\set{\ContSysDT{\Sigma_i}}_{i\in I}$ and $\set{\ContSysA{i}{\Sigma_i}}_{i\in I}$ of $\set{\Sigma_i}_{i\in I}$, s.t. for all $i\in I$, $\ContSysDT{\Sigma_i}$ and $\ContSysA{i}{\Sigma_i}$ are disturbance bisimilar with parameters $(\varepsilon_i, \tilde{\varepsilon}_i)$. Then we adapt our result from \cite{majumdar2016compositional}, and prove that the isomorphic networks of $\set{\ContSysDT{\Sigma_i}}_{i\in I}$ and $\set{\ContSysA{i}{\Sigma_i}}_{i\in I}$, which are isomorphic to the network of $\set{\Sigma_i}_{i\in I}$ as well, are again disturbance bisimilar.

\subsection{Network of Stochastic Control Systems}

We first formalize networks of stochastic control systems and their abstractions by locally treating state trajectories of neighboring systems as disturbances.

Let $I$ be an index set (e.g., $I = \set{1,\ldots, N}$ for some natural number $N$) and let
$\IntCon\subseteq I\times I$ be a binary irreflexive \emph{connectivity relation} on $I$. 
Furthermore, let $I'\subseteq I$ be a subset of systems with $\IntCon' := (I'\times I')\cap \IntCon$.
For $i\in I$ we define 
$\nbr_{\mathcal{I}}(i) = \set{j\mid (j,i)\in \IntCon}$ and extend this notion to subsets of systems $I'\subseteq I$ as $\nbr_{\mathcal{I}}(I') = \set{j\mid \exists i\in I'.j\in \nbr_{\mathcal{I}\setminus\mathcal{I}'}(i)}$. 

Intuitively, a set of systems can be imagined to be the set of vertices $\set{1,2,\ldots,|I|}$ of a directed graph $\mathcal{G}$, and $\mathcal{I}$ to be the corresponding adjacency relation. Given any vertex $i$ of $\mathcal{G}$, the set of incoming (resp. outgoing) edges are the inputs (resp. outputs) of a subsystem $i$, and $\nbr_{\mathcal{I}}(i)$ is the set of neighboring vertices from which the incoming edges originate. 

Let $\Sigma_i = (X_i, U_i, \mathcal{U}_i, W_i, \mathcal{W}_i, f_i,\sigma_i)$,
for $i\in I$, be a collection of stochastic control systems.
We say that the set of stochastic control systems $\set{\Sigma_i}_{i\in I}$ is \emph{compatible for composition} w.r.t.\ the interconnection relation $\mathcal{I}$, if 
for each $i\in I$, we have
$W_i = \prod_{j\in \nbr_\mathcal{I}(i)} {X_j}$. By slightly abusing notation we write $w_i=\prod_{j\in \nbr_\mathcal{I}(i)} \set{x_j}$ for $x_j\in X_j$ and $w_i\in W_i$ as a short form for the single element of the set $\prod_{j\in \nbr_\mathcal{I}(i)} \set{x_j}$. We extend this notation to all sets with a single element.

 Let $I'\subset I$ be a subset of systems in the network. We divide the set of disturbances $W_i$ for any $i\in I'$ into the sets of coupling and external disturbances, defined by
 $W_i^c = \prod_{j\in \nbr_{\mathcal{I}'}(i)} {X_j}$ and $W_i^e = \prod_{j\in\nbr_{\IntCon\setminus\IntCon'}(i)} {X_j}$, respectively.

If $\set{\Sigma_i}_{i\in I}$ is compatible, we define the \emph{composition} of any subset $I'\subseteq I$ of systems as the stochastic control system
$\llbracket \Sigma_i\rrbracket_{i\in I'}= (X, U, \mathcal{U}, W, \mathcal{W}, f,\sigma)$ where $X$, $U$ and $W$ are defined as
$X = \prod_{i\in I'} {X_i}$,
$U = \prod_{i\in I'} {U_i}$, and
$W = \prod_{j\in \nbr_{\mathcal{I}}(I')} {X_j}$.
Furthermore, $\mathcal{U}$ and $\mathcal{W}$ are defined as the sets of functions $\mu : \poszreal{} \rightarrow U$ and 
$\nu : \Omega \times \poszreal{} \rightarrow W$
such that the projection $\mu_i$ of $\mu$ on to $U_i$ (written $\mu_i=\mu|_{U_i}$) belongs to $\mathcal{U}_i$, and the projection $\nu_i^e$ of $\nu$ on to $W_i^e$ belongs to $\mathcal{W}_i^e$.
The composed drift is then defined as $f(\prod_{i\in I'} \set{x_i}, \prod_{i\in I'} \set{u_i}, \prod_{i\in I'} \set{w_i^e}) = \prod_{i\in I'} \set{f_i(x_i, u_i, w_i^c \times w_i^e)}$, where $w_i^c = \prod_{j\in \nbr_{\mathcal{I}'}(i)} \set{x_j}$, and the composed diffusion is defined as $\sigma(\prod_{i\in I'} \set{x_i}) = Diag\left( \sigma_1(x_1),\ldots, \sigma_{|I|}(x_{|I|}) \right)$. 
The Brownian motion of the overall system is defined as: $dB_t = \begin{bmatrix} dB_{1,t}& \hdots& dB_{|I|,t}\end{bmatrix}^T$. 

If $I'=I$, then $\Sigma$ is undisturbed, modeled by $W:=\set{0}$.
It is easy to see that $\llbracket \Sigma_i\rrbracket_{i\in I'}$ is again a stochastic control system in the sense of \Def~\ref{Def_control_sys}.
Networks of discrete time stochastic metric systems 
($\ContSysDT{\Sigma_i}$)
and of abstract metric systems
($\ContSysA{i}{\Sigma_i}$)
are defined analogously.

\begin{remark}
	Note that we assume a nice structure of the network: the diffusion functions and the Brownian motions of the systems in a network are decoupled from the states of the other systems.
	This is explicitly induced via the SDE \eqref{eq0} as the diffusion terms $\sigma_i(\cdot)$ are only functions of system's state and not the disturbance.
	However since the states of the systems are coupled through the drift functions, the respective random variables are implicitly dependent.
\end{remark}

\subsection{Simultaneous Approximation}
\label{sec:CA:Assumptions}
Given $I$ and $I' \subseteq I$, consider a set of compatible stochastic control systems $\set{\Sigma_i}_{i\in I}$, the subset composition $\llbracket \Sigma_i\rrbracket_{i\in I'}= (X, U, \mathcal{U}, W, \mathcal{W}, f,\sigma)$ and 
a global time-sampling parameter $\tau$. 
Then we can apply \Def~\ref{def:control-sys-disctime} and \Def~\ref{def:abstract system} to each $\Sigma_i$ to construct the corresponding 
metric systems $\ContSysDT{\Sigma_i}$ and $\ContSysAi{\Sigma_i}$. To be able to do that, we need to equip $W_i$ with a vector-valued metric $\vmetric_i:W_i\times W_i\rightarrow\real{|\nbr_{\mathcal{I}}(i)|}_{\geq 0}$ and define $\Wt_i$ for all $i\in I$ s.t. Ass.~\ref{ass:Wt} holds. 
Intuitively, $\vmetric_i(w_i,w_i')$ is a vector with dimension $|\nbr_{\mathcal{I}}(i)|$, where the $j^{\text{th}}$ entry measures the mismatch of the respective state vector of the $j^{\text{th}}$ neighbor of $i$. 
We define $\Wt_i$ as the product of state spaces of $\ContSysA{j}{\Sigma_j}$, i.e., the abstraction of its neighbors,
\begin{equation}\label{equ:Wti}
\textstyle \Wt_i:=\prod_{j \in \nbr_{\mathcal{I}}(i)} [X_j]_{\eta_j}.
\end{equation}

\begin{lemma}
	\label{lem:Wt}
 Let $\Sigma_i = (X_i, U_i, \mathcal{U}_i, W_i, \mathcal{W}_i, f_i,\sigma_i)$,$i\in I$, be a set of compatible stochastic control systems and the set of abstract metric systems
 $\ContSysA{i}{\Sigma_i} = ([X_i]_{\eta_i},\DiscSpace{U}{\IPQnt{i}},\IPECurveA{i}, \Wt_i,\IPICurveA{i},\deltaA{i})$ are constructed according to \Def~\ref{def:abstract system}, where $W_i$ is equipped with metric \eqref{equ:defvmetric_composed} and $\Wt_i$ as defined in \eqref{equ:Wti}.
Select local quantization parameters $\set{\StateQnt{i}}_{i\in I}$ \st $\eta_i\leq\varepsilon_i$. Then Ass.~\ref{ass:Wt} holds for every $i\in I$  with $\tilde{\varepsilon}_i$ defined as
\begin{equation}
\label{equ:tildevarepsilon 1}
\textstyle \tilde{\varepsilon}_i:=\prod_{j\in\nbr_{\mathcal{I}}(i)} \set{\varepsilon_j}.
\end{equation}
\end{lemma}

Given \Lem~\ref{lem:Wt}, it immediately follows that the sets $\set{\ContSysDT{\Sigma_i}}_{i\in I'}$ and $\set{\ContSysAi{\Sigma_i}}_{i\in I'}$ of metric systems are again compatible. 

In order to guarantee the result of \Thm~\ref{thm:sim-approx-single} for the network, we have additionally used \Ass~\ref{ass:psi} which essentially bounds the effect of the disturbances on the state evolution. Given the particular choice of disturbances in the network as state trajectories of neighboring systems, we can explicitly compute function $\psi(t)$ in \Ass~\ref{ass:psi} using the following proposition from \cite[\Thm~4.3]{MAO_stochastic}.

\begin{proposition}
	\label{prop:bound on moment}
	Under \Ass~\ref{ass:linear_growth} the solution process $\xi_{a\mu\nu}(\cdot)$ satisfies the inequality
	\begin{equation*}
		\mathbb E\left[\|\xi_{a\mu\nu}(t)-\xi_{a\mu\nu}(s)\|_2^2\right]\le C|t-s|,\quad\forall s,t\in[0,\tau],
	\end{equation*}
for any $\tau>0$, where $\|\cdot\|_2$ indicates the 2-norm of a vector.
The constant $C := 2\left(1+\mathbb E\|a\|_2^2\right)(\tau+1)e^{\alpha \tau}$ with $\alpha:=K+2\sqrt{K}$ and $K$ from \Ass~\ref{ass:linear_growth}.
\end{proposition}

The next lemma follows  from \Prop~\ref{prop:bound on moment}.
\begin{lemma}
	\label{lem:bound_moment}
	Given a set of stochastic control systems $\set{\Sigma_i}_{i\in I}$ which is compatible for composition, let each system $\Sigma_i$ satisfy \Ass~\ref{ass:linear_growth} with constant $K_i$ in \eqref{eq:lin_growth}. Then \Ass~\ref{ass:psi} holds for each $\Sigma_i$ with $\KInf$ function
	\begin{equation*}
		\textstyle \psi_i(t):=
		\left[t(t+1)\sum_{j\in\nbr_\mathcal{I}(i)}\beta_j e^{\alpha_j t}\right]^\frac{1}{2},
	\end{equation*}
where $\alpha_j:=K_j+2\sqrt{K_j}$ and 
	$\beta_j = 2\left(1+ \sup_{x_j\in X_j}\|x_j\|_2^2\right)$.
\end{lemma}

Lemmas~\ref{lem:Wt}-\ref{lem:bound_moment} show that the assumptions of \Sec~\ref{sec:Dist_bisimulation} on disturbance sets of $\Sigma_i$ hold after composition.
Then the next theorem follows from \Thm~\ref{thm:sim-approx-single} which establishes simultaneous disturbance bisimilarity between abstractions of components in a network.
In this theorem, using the results in \Thm~\ref{thm:sim-approx-single}, we give conditions on all local state, input, and 
disturbance quantization parameters in a composed stochastic control system $\llbracket \Sigma_i\rrbracket_{i\in I}$ which allow for a simultaneous construction of local abstractions $\ContSysAi{\Sigma_i}$ using \Def~\ref{def:abstract system} 
such that they are disturbance bisimilar with parameters $(\varepsilon_i,\tilde{\varepsilon}_i)$ 
to their respective discrete-time stochastic metric systems $\ContSysDT{\Sigma_i}$.

\begin{theorem}
	\label{thm:sim-appx}
Let $\set{\Sigma_i}_{i\in I}$ be a set of compatible stochastic control systems, each admitting a 
$\delta$-ISS-$M_2$ Lyapunov function 
	${\Lypv{i}}$ witnessed by $\kappa_i$, $\alphalow{i}$, $\alphahigh{i}$, $\sigmau{i}$, and $\sigmad{i}$, and let
	$\gammafunc{i}$ be a $\KInf$ function \st \eqref{supplement} holds.
Let $\set{\ContSysDT{\Sigma_i}}_{i\in I}$ be the set of discrete-time stochastic metric systems 
induced by $\set{\Sigma_i}_{i\in I}$ and let $\set{\ContSysAi{\Sigma_i}}_{i\in I}$ be the set of countable deterministic metric systems
  induced by $\set{\Sigma_i}_{i\in I}$ and $\Wt_i$ as in \eqref{equ:Wti}.
If all local quantization parameters $\set{\StateQnt{i},\varepsilon_i,\IPQnt{i}}_{i\in I}$ simultaneously fulfill $\StateQnt{i}\le \varepsilon_i$ and
		\begin{multline}
	0\leq \StateQnt{i} \leq \textrm{min}\Big\lbrace(\alphahigh{i})^{-1}\circ\alphalow{i}(\varepsilon_i^2),\gammafunc{i}^{-1}\big[ (1-\mathsf{e}^{-\kappa_i\tau})\alphalow{i}(\varepsilon_i^2)\\
	- \frac{1}{\mathsf{e}\kappa_i}\sigma_{u,i}(\omega_i) - \frac{1}{\mathsf{e}\kappa_i}\sigma_{d,i}(\psi_i(\tau) + \norm{\tilde{\varepsilon}_i}) \big] - (h_i(\sigma_i,\tau))^{\frac{1}{2}}\Big\rbrace,
	\label{eq:condition on StateQnt i}
	\end{multline}
	with $\set{\tilde{\varepsilon}_i}_{i\in I}$ defined as \eqref{equ:tildevarepsilon 1}, then the relation
	\begin{align*}
	\eqclass{i} = &\left\lbrace (\hat{x}_i,x_i)\in[X_i]_{\eta_i}\times X_{i,\mathsf r} \ |\ \mathbb{E}[\Lypv{i}(\hat{x}_i,x_i)]\leq\alphalow{i}(\varepsilon_{i}^2) \right\rbrace
	\end{align*}
	is a disturbance bisimulation relation in the second moment with parameters $(\varepsilon_i,\tilde{\varepsilon}_i)$ between
	  $\ContSysA{i}{\ContSysSymb{i}}$ and $\ContSysDT{\ContSysSymb{i}}$ for all $i\in I$.
\end{theorem}

 \subsection{Composition of Approximations}
 We have discussed in \Sec~\ref{sec:CA:Assumptions} that the sets $\set{\ContSysDT{\Sigma_i}}_{i\in I}$ and $\set{\ContSysAi{\Sigma_i}}_{i\in I}$ of metric systems is compatible.
 We also established conditions on local quantization parameters under which the metric systems $\ContSysDT{\Sigma_i}$ and $\ContSysAi{\Sigma_i}$ are disturbance bisimilar for any $i\in I$.
 
 We now use the fundamental property of disturbance bisimulation relation proved in \cite{majumdar2016compositional} that disturbance bisimilarity is preserved under composition of components in a network.
 This property together with \Thm~\ref{thm:sim-appx} result in the following theorem that explicitly gives the disturbance bisimulation relation on the composed abstractions of components in a network.

\begin{theorem}\label{cor:comp-appx}
		Given the preliminaries of \Thm~\ref{thm:sim-appx} and $I'\subseteq I$, let $\llbracket\ContSysDT{\Sigma_i}\rrbracket_{i\in I'}$ and $\llbracket\ContSysAi{\Sigma_i}\rrbracket_{i\in I'}$
		be systems with state spaces $X_{\mathsf r}$ and $\StateAb{}$, composed from the sets $\set{\ContSysDT{\Sigma_i}}_{i\in I}$ and $\set{\ContSysAi{\Sigma_i}}_{i\in I}$, respectively. Then the relation
			\begin{align}\label{equ:eqclass}
				\eqclass{} = &\lbrace (\transpose{\hat{q}^T_{1}\delimit\ldots\delimit\hat{q}^T_{|I'|}},\transpose{ q^T_{1}\delimit\ldots\delimit q^T_{|I'|}})\in \StateAb{}\times X_{\mathsf r} \ |\notag\\
				&\quad(\hat{q}_i,q_i)\in\eqclass{i}, \forall i\in I'
				)\rbrace
			\end{align} 
		 is a disturbance bisimulation relation between $\llbracket\ContSysDT{\Sigma_i}\rrbracket_{i\in I'}$ and $\llbracket\ContSysAi{\Sigma_i}\rrbracket_{i\in I'}$ with parameters 
		 		\begin{align*}
				\textstyle\varepsilon{} \textstyle= \norm{\prod_{i\in I'}\set{\varepsilon_i}} \text{ and }~ 
				\tilde{\varepsilon}\textstyle=\prod_{j \in \nbr_{\mathcal{I}}(I')} \set{\varepsilon_j}.
			\end{align*}
	\end{theorem}

Note that in the special case $I'=I$ the composed system replaces the overall network without extra external disturbances. In this case it is easy to see that the relation in \Thm~\ref{cor:comp-appx} simplifies to a usual bisimulation relation.

\begin{corollary}
 Given the premises of \Thm~\ref{cor:comp-appx} and that $I'=I$, the relation $\eqclass{}$ in \eqref{equ:eqclass} is an $\varepsilon$-approximate bisimulation relation between $\llbracket\ContSysDT{\Sigma_i}\rrbracket_{i\in I}$ and $\llbracket\ContSysAi{\Sigma_i}\rrbracket_{i\in I}$.
\end{corollary}

\section{Conclusion}
In this paper, we extended our previous result on compositional abstraction based control for non-probabilistic control systems to stochastic control systems. We gave sufficient conditions s.t. a stochastic control system, admitting a $\delta$-ISS-$M_q$ Lyapunov function and subjected to small mismatch in the continuous and abstract disturbances, admits a disturbance bisimilar abstract system. Then we used the property of disturbance bisimulation to show that given a network of stochastic control systems, the abstract systems can be computed compositionally. One can then use this paper's claim for compositional synthesis of controllers for networks of stochastic control systems, as is done in \cite[\Sec~VII]{majumdar2016compositional} for network of deterministic systems.

\bibliographystyle{plain}

\begin{appendix}

\subsection{Proof of \Thm~\ref{thm:sim-approx-single}}
\newcounter{eqn}

	First observe that $\StateA{}\subset\StateDT{}$, hence the metric $d$ on $\StateDT{}$ is also a metric on $\StateA{}$. Now we prove the three parts of \Def~\ref{def:DisturbanceBisimulation} separately.\\
	\begin{inparaenum}[(a)]
		\item By definition of $\eqclass{}$ in \eqref{eqn:local simulation relation}, $(\hat{x},x)\in \eqclass{}$ implies
		\begin{align}
			\stepcounter{eqn}
			\tag{\Alph{subsection}.\arabic{eqn}} 
			\label{convexity}
			d(\hat{x},x) = \left(\mathbb{E}\left[ \norm{\hat{x}-x}^2\right]\right)^{1/2}\leq
			\left(\underline\alpha^{-1}\left(\mathbb{E}\left[V(\hat{x},x)\right]\right)\right)^{1/2}
			\leq\varepsilon.
		\end{align}
		We used the convexity assumption of $\underline\alpha$ and the Jensen inequality \cite{oksendal} to show the inequalities in (\ref{convexity}). \\
		\item Given a pair $(\hat{x}, x)\in\eqclass{}$, for any $\mu\in\IPECurveDT{}$, observe that there exists a $\hat{\mu} \in \IPECurveA{}$ s.t.\ $\norm{u_{\hat{\mu}}-u_\mu}\leq \IPQnt{}$ holds.
		Given any $\hat{\nu}\in\IPICurveA{}$ and $\nu\in\IPICurveDT{}$ s.t. $\vmetric(w_{\hat{\nu}},w_\nu)\leq \DQnt{}$ holds, observe that $\norm{w_{\hat{\nu}}-w_\nu}\leq\norm{\vmetric(w_{\hat{\nu}},w_\nu)}\leq\norm{\DQnt{}}$ from \eqref{equ:Vtildeeps}. Now we can apply transitions 
		$\delta_\tau (x,\mu,\nu) = \xi_{x\mu\nu}(\tau) = x'$, $\xi_{\hat{x}\hat{\mu}\hat{\nu}}(\tau) = z$, $\overline{\xi}_{\hat{x}\hat{\mu}\hat{\nu}}(\tau) = \overline{z}$, and observe that there exists a $\hat{x}' \in \StateA{}$ s.t. $\norm{\hat{x}'-\overline{z}} \leq \eta$, and hence we have $\delta_{\tau\eta\omega}(\hat{x},\hat{\mu},\hat{\nu}) = \hat{x}'$.
	   	Now consider the following derivation:
		\begin{small}
		\begin{align} 
		\stepcounter{eqn}
		\tag{\Alph{subsection}.\arabic{eqn}}
		\label{equ:proof:derivation1}
		&\mathbb{E}\left[V(\hat{x}',x')\right]\\ \notag
		&=\mathbb{E}\left[V(z,x')+V(\hat{x}',x')-V(z,x')\right]\\ \notag
		&=  \mathbb{E}\left[V(z,x')\right]+\mathbb{E}\left[V(\hat{x}',x')-V(z,x')\right]\\\notag
		&\leq\underline\alpha(\varepsilon^2)\mathsf{e}^{-\kappa\tau}+\frac{1}{\mathsf{e}\kappa}\sigma_u(\omega) + \frac{1}{\mathsf{e}\kappa}\sigma_d(\psi_i(\tau) + \norm{\tilde{\varepsilon}})\\ \notag
		&\qquad +\mathbb{E}\left[\widehat\gamma\left( \norm{\hat{x}'-z} \right)\right]\\\notag
		&\leq\underline\alpha(\varepsilon^2)\mathsf{e}^{-\kappa\tau} +\frac{1}{\mathsf{e}\kappa}\sigma_u(\omega) + \frac{1}{\mathsf{e}\kappa}\sigma_d(\psi_i(\tau) + \norm{\tilde{\varepsilon}})\\\notag
		&\qquad +\widehat\gamma\left(\mathbb{E}\left[ \norm{\hat{x}' -\overline{z} + \overline{z} -z} \right]\right)\\\notag
		&\leq\underline\alpha(\varepsilon^2)\mathsf{e}^{-\kappa\tau}+ \frac{1}{\mathsf{e}\kappa}\sigma_u(\omega) + \frac{1}{\mathsf{e}\kappa}\sigma_d(\psi_i(\tau) + \norm{\tilde{\varepsilon}})\\\notag
		&\qquad+\widehat\gamma\left(\mathbb{E}\left[ \norm{\overline{z}-z}\right]+\norm{\hat{x}'-\overline{z}}\right)\\\notag 
		&\leq\underline\alpha(\varepsilon^2)\mathsf{e}^{-\kappa\tau}+ \frac{1}{\mathsf{e}\kappa}\sigma_u(\omega) + \frac{1}{\mathsf{e}\kappa}\sigma_d(\psi_i(\tau) + \norm{\tilde{\varepsilon}})\\ \notag
		&\qquad +\widehat\gamma\left((h(\sigma,\tau))^{\frac{1}{2}}+\eta\right)\\\notag
		&\leq\underline\alpha(\varepsilon^2).\notag
		\end{align}
		\end{small}
	   	Hence by \Eqn~\eqref{eqn:local simulation relation}, $(\hat{x}',x')\in\eqclass{}$.\\
		\item Given a pair $(\hat{x}, x)\in\eqclass{}$, for any $\hat{\mu}\in\IPECurveA{}$, observe that we can choose $\mu \in \IPECurveDT{}$ s.t. $\mu = \hat{\mu}$, i.e., $\norm{u_{\hat{\mu}}-u_\mu} = 0$.    		
   		Given any $\nu\in\IPICurveDT{}$ and $\hat{\nu}\in\IPICurveA{}$ s.t.\ $\vmetric(w_{\hat{\nu}},w_\nu)\leq \DQnt{}$, we have as before $\delta_\tau (x,\mu,\nu) = \xi_{x\mu\nu}(\tau) = x'$, $\xi_{\hat{x}\hat{\mu}\hat{\nu}}(\tau) = z$, $\overline{\xi}_{\hat{x}\hat{\mu}\hat{\nu}}(\tau) = \overline{z}$, and observe that there exists a $\hat{x}' \in \StateA{}$ s.t. $\norm{\hat{x}'-\overline{z}} \leq \eta$, and hence we have $\delta_{\tau\eta\omega}(\hat{x},\hat{\mu},\hat{\nu}) = \hat{x}'$.
		With a very similar derivation as in \eqref{equ:proof:derivation1} it follows from \Eqn~\eqref{eqn:local simulation relation} that $(\hat{x}',x')\in\eqclass{}$.

	\end{inparaenum}

\subsection{Proof of other statements}
\begin{proof}[\Lem \ref{lem:lyapunov}]
	The proof of Lemma \ref{lem:lyapunov} can be obtained from the proof of \Lem~3.4 in \cite{zamani2014symbolic} by replacing all instances of $f(x,u)$, $f(z,u)$, $f(x',u)$ and $f(x',u')$ with $f(x,u,w)$, $f(z,u,w)$ , $f(x',u,w)$ and $f(x'.u',w')$ respectively, and defining the positive constant $\kappa = \widetilde{\kappa}$, the $\KInf$ functions $\sigma_u(r) = \left( nL_u^2/\widetilde{\kappa} \right)\norm{\sqrt{P}}^2r^2$ and $\sigma_d(r) = \left( nL_w^2/\widetilde{\kappa} \right)\norm{\sqrt{P}}^2r$.
\end{proof}

\begin{proof}[\Lem~\ref{lem:moment est}]
	The proof of Lemma \ref{lem:moment est} follows closely the proof of \Lem~3.8 in \cite{zamani2014symbolic} and hence is omitted.
\end{proof}

\begin{proof}[\Lem~\ref{lem:Wt}]
	Pick any $i\in I$, $w_i\in W_i$ and observe that $w_i=\prod_{j \in \nbr_{\mathcal{I}}(i)} \set{x_{j}}$. By the choice of $\StateA{j}$ as $\DiscSpace{X_j}{\StateQnt{j}}$ we furthermore know that for any $x_j$ there exists $\hat{x}_j$ \st $\norm{x_j-\hat{x}_j}\leq\eta_j\leq\varepsilon_j$. Now recall that  $\Wt_i=\prod_{j \in \nbr_{\mathcal{I}}(i)} \StateA{j}=\prod_{j \in \nbr_{\mathcal{I}}(i)} \DiscSpace{X_j}{\StateQnt{j}}$. 
	Using the definition of $\tilde{\varepsilon}_i$ in \eqref{equ:tildevarepsilon 1} and $\vmetric$ in \eqref{equ:defvmetric_composed} we therefore know that for any $w_i\in W_{i}$ there exists $\tilde{w}_i\in \Wt_i$ s.t. $\vmetric(w_i,\wt_i)=\prod_{j \in \nbr_{\mathcal{I}}(i)} \set{\norm{x_j - \hat{x}_j}} \leq\prod_{j \in \nbr_{\mathcal{I}}(i)} \set{\varepsilon_j}=\tilde{\varepsilon}_i$. 
	Furthermore, $\norm{w_i-\wt_i} = \norm{\prod_{j\in\nbr_{\mathcal{I}}(i)}\set{x_j - \hat{x}_j}} = \norm{\prod_{j\in\nbr_{\mathcal{I}}(i)}\set{\norm{x_j - \hat{x}_j}}} = \norm{\vmetric(w_i,\wt_i)}$. 
\end{proof}

\begin{proof}[\Lem~\ref{lem:bound_moment}]
	The proof follows from the following derivation:
	\begin{align*}
	&d(\nu_i(t),\hat{\nu}_i(t))\\
	= &\left(\EE\left[ \norm{\nu_i(t) - \hat{\nu}_i(t)}^2\right]\right)^\frac{1}{2}\\
	= &\left(\EE\left[ \norm{\nu_i(t) - \hat{\nu}_i(0)}^2 \right]\right)^\frac{1}{2}\\
	= &\left( \EE\left[ \norm{\nu_i(t) - \nu_i(0) + \nu_i(0) -\hat{\nu}_i(0)}^2 \right] \right)^\frac{1}{2}\\
	\leq &\left( \EE\left[ \norm{\nu_i(t) - \nu_i(0)}^2\right] \right)^\frac{1}{2} + \left( \EE\left[ \norm{\nu_i(0) -\hat{\nu}_i(0)}^2 \right] \right)^\frac{1}{2}\\
	= &\left( \EE\left[ \norm{\prod_{j\in\nbr_\mathcal{I}(i)}\set{\xi_j(t) - \xi_j(0)}}^2\right] \right)^\frac{1}{2} + \norm{\tilde{\varepsilon}_i}\\
	= &\left( \EE\left[ \left(\sup_{\substack{j\in\nbr_\mathcal{I}(i)}}
	\norm{\xi_j(t) - \xi_j(0)}\right)^2\right] \right)^\frac{1}{2} + \norm{\tilde{\varepsilon}_i}\\
	\leq &\left( \EE\left[ \sum_{\substack{j\in\nbr_\mathcal{I}(i)}} \norm{\xi_j(t) - \xi_j(0)}^2\right] \right)^\frac{1}{2} + \norm{\tilde{\varepsilon}_i}\\
	\le  &\left( \sum_{j\in\nbr_\mathcal{I}(i)} \EE\left[  \norm{\xi_j(t) - \xi_j(0)}_2^2\right] \right)^\frac{1}{2} + \norm{\tilde{\varepsilon}_i}\\
	\leq &\left(\sum_{j\in\nbr_\mathcal{I}(i)}2t\left(1+\mathbb E\|\xi_j(0)\|_2^2\right)(t+1)e^{\alpha_j t}\right)^\frac{1}{2} + \norm{\tilde{\varepsilon}_i}\\
	\leq &\left(\sum_{j\in\nbr_\mathcal{I}(i)}2t\left(1+ \sup_{x_j\in X_j}\|x_j\|_2^2\right)(t+1)e^{\alpha_j t}\right)^\frac{1}{2} + \norm{\tilde{\varepsilon}_i}
	\end{align*}
	where $\alpha_j = K_j+2\sqrt{K_j}$ and $K_j$ is the constant $K$ as given in \Prop~\ref{prop:bound on moment} for the $j$-th system, and the last step follows from \Prop~\ref{prop:bound on moment}. We define for system $\Sigma_i$ the $\KInf$ function
	 $\psi_i(t):= \left[t(t+1)\sum_{j\in\nbr_\mathcal{I}(i)}\beta_j e^{\alpha_j t}\right]^\frac{1}{2}$, which concludes the proof.
\end{proof}

\end{appendix}

\end{document}